\documentclass[lettersize,journal]{IEEEtran}
\usepackage{amsmath,amsfonts}
\usepackage{algorithmic}
\usepackage{algorithm}
\usepackage{array}
\usepackage[caption=false,font=normalsize,labelfont=sf,textfont=sf]{subfig}
\usepackage{textcomp}
\usepackage{stfloats}
\usepackage{url}
\usepackage{verbatim}
\usepackage{graphicx}
\usepackage{cite}
\begin{document}

\title{VLM- and LLM-Driven Multi-Agent System for\\ PET Image Denoising}

\author{Boxiao Yu, Savas Ozdemir, Yang Xing, Fumio Hashimoto, Jiong Wu, Yizhou Chen, Axel Rominger, Ruogu Fang, Kuangyu Shi, Tinsu Pan, Kuang Gong
\thanks{This work was supported by NIH grants R01EB034692 and R01AG078250. (Corresponding author: Kuang Gong)}
\thanks{This work did not involve human subjects or animals in its research.}
\thanks{Boxiao Yu, Yang Xing, Fumio Hashimoto, Jiong Wu, Ruogu Fang, and Kuang Gong are with the J. Crayton Pruitt Family Department of Biomedical Engineering, University of Florida, Gainesville, FL, 32611, USA. (e-mails: boxiao.yu@ufl.edu, yxing2@ufl.edu, fumio.hashimo@ufl.edu, wujiong@ufl.edu, Ruogu.Fang@bme.ufl.edu, and KGong@bme.ufl.edu).}
\thanks{Savas Ozdemir is with the Department of Radiology, University of Florida, Jacksonville, FL, 32209, USA. (e-mail: savas.ozdemir@jax.ufl.edu).}
\thanks{Yizhou Chen, Axel Rominger, and Kuangyu Shi are with the Department of Nuclear Medicine, University of Bern, Bern, Switzerland. (e-mail: yizhou.chen@students.unibe.ch; e-mail: axel.rominger@insel.ch; e-mail: kuangyu.shi@unibe.ch).}
\thanks{Tinsu Pan is with the Department of Imaging Physics, University of Texas MD Anderson Cancer Center, Houston, TX, 77230, USA. (e-mail: tpan@mdanderson.org).}}

\markboth{IEEE Transactions on Radiation and Plasma Medical Sciences,~Vol.~XX, No.~X, August~2026}%
{Shell \MakeLowercase{\textit{et al.}}: A Sample Article Using IEEEtran.cls for IEEE Journals}


\maketitle

\begin{abstract}
Positron emission tomography (PET) imaging suffers from limited spatial resolution and low signal-to-noise ratio, which can compromise quantitative accuracy and lesion detectability. Deep learning–based denoising methods have demonstrated strong potential for improving PET image quality. However, their practical deployment in real-world settings remains challenging, often requiring multiple specialized models and expert interventions, such as identifying motion-induced misregistration artifacts, estimating noise levels to select an appropriate denoiser, and performing lesion-focused quantitative assessment after denoising. Recent advances in vision–language models (VLMs) for image quality understanding and large language models (LLMs) for contextual reasoning provide new opportunities for automated, decision-driven workflows. Inspired by expert workflows for PET image quality enhancement, we propose an VLM– and LLM–driven multi-agent PET denoising framework that dynamically assesses image quality and lesion status, autonomously selects optimal denoising models and parameters, and enables closed-loop feedback with rollback mechanisms. Experiments were conducted on Siemens Biograph Vision Quadra PET/CT data with $1/20$ and $1/50$ low-dose settings. Individual module evaluations demonstrated the reliability of the agentic components, while the complete framework achieved higher PSNR and SSIM than UNet, GAN, and DDPM baselines at both dose levels. These preliminary results demonstrate the feasibility of using a closed-loop multi-agent framework to adapt PET denoising strategies to different image conditions.
\end{abstract}


\section{Introduction}
\IEEEPARstart{P}{ositron} emission tomography (PET) is a molecular imaging modality that enables noninvasive, quantitative assessment of disease across oncology, neurology, and cardiology~\cite{ming2020progress,barthel2015pet,dorbala2014cardiac}. However, PET image quality is constrained by limited photon coincidences and various physical degradation factors~\cite{gong2017sinogram}. Additional reductions in injected dose or acquisition time, while attractive for lowering radiation exposure and increasing scanner throughput, can further degrade the signal-to-noise ratio and may bias standardized uptake values (SUVs), particularly in small or low-contrast lesions. A further challenge is that real-world scans span a wide spectrum of dose/count levels, patient sizes, and reconstruction settings. Consequently, a denoiser that performs well under one dose condition may fail under another~\cite{liu2024cross}, leading to either insufficient noise suppression or excessive smoothing that compromises lesion detectability. Therefore, dose-adaptive PET denoising with reliable validation is essential for robust clinical translation.

Deep learning has substantially advanced PET image denoising, spanning convolutional networks~\cite{jaudet2021impact,cui2019pet,kaplan2019full,nerella2023transformers,hashimoto2019dynamic,gong2018pet,zhou2021mdpet,li2026deep}, adversarial learning~\cite{wang20183d,fu2023aigan,song2020pet,lei2019whole,zhou2020supervised}, and more recent generative models~\cite{yu2024pet,yu2025robust,jiang2023pet,shen2023pet,webber2025supervised,xie2023ddpet,pan2024full,lin2025multimodal,shen2024bidirectional,sun2025legopet}. While these approaches can produce visually appealing images and improve common fidelity metrics under controlled experimental conditions, their robustness in heterogeneous clinical settings remains limited. As a result, deploying these models in real-world workflows often involves multiple manual steps requiring domain expertise, such as checking for motion-induced PET/CT misregistration, assessing noise severity to determine the appropriate model, selecting inference parameters to balance noise and contrast, and performing post-denoising lesion-focused evaluation to ensure lesions are not smoothed out. Moreover, many existing denoising methods are optimized for specific dose levels, scanner protocols, or preprocessing pipelines. 
These limitations make it difficult to construct a unified and automated denoising workflow.

A straightforward solution would be to encode expert decisions using predefined rules. However, PET denoising decisions often depend on several interrelated factors, including dose/count condition, organ noise, lesion status, and the characteristics of the available denoising models. A large rule table could be manually constructed for a fixed set of conditions, but it would require substantial revision whenever a new dose level, denoiser, or adjustable parameter is introduced. More importantly, expert denoising is not always a one-step decision. When an initial strategy produces insufficient noise reduction or alters lesion uptake, the next plan must depend on what was attempted and how it failed. Such history-dependent adaptation requires task-specific experience and contextual reasoning, which are difficult to represent using static decision rules.

Recent advances in vision-language models (VLMs) and large language models (LLMs) provide a potential solution to this problem. VLMs have demonstrated strong capability in joint image–text understanding~\cite{radford2021learning, zhang2023biomedclip, li2023llava}, enabling tasks such as quality grading, artifact recognition, and attribute classification via  prompts~\cite{you2024depicting,wu2023q,wu2024q, cao2024synartifact}. LLMs provide complementary strengths in contextual reasoning~\cite{wei2022chain,yao2022react}, external knowledge retrieval~\cite{lewis2020retrieval, zakka2024almanac}, and tool coordination~\cite{schick2023toolformer, shen2023hugginggpt}, allowing heterogeneous information to be combined into coherent decision strategies and revise decisions according to previous outcomes. Together, these capabilities make it possible to move beyond a fixed image-to-image denoising pipeline toward an intelligent agentic workflow that can perceive scan conditions, select appropriate strategies, execute specialized tools, and verify outcomes in a closed loop, analogous to expert-driven PET image processing.


\begin{figure}
\centering
\includegraphics[width=0.5\textwidth]{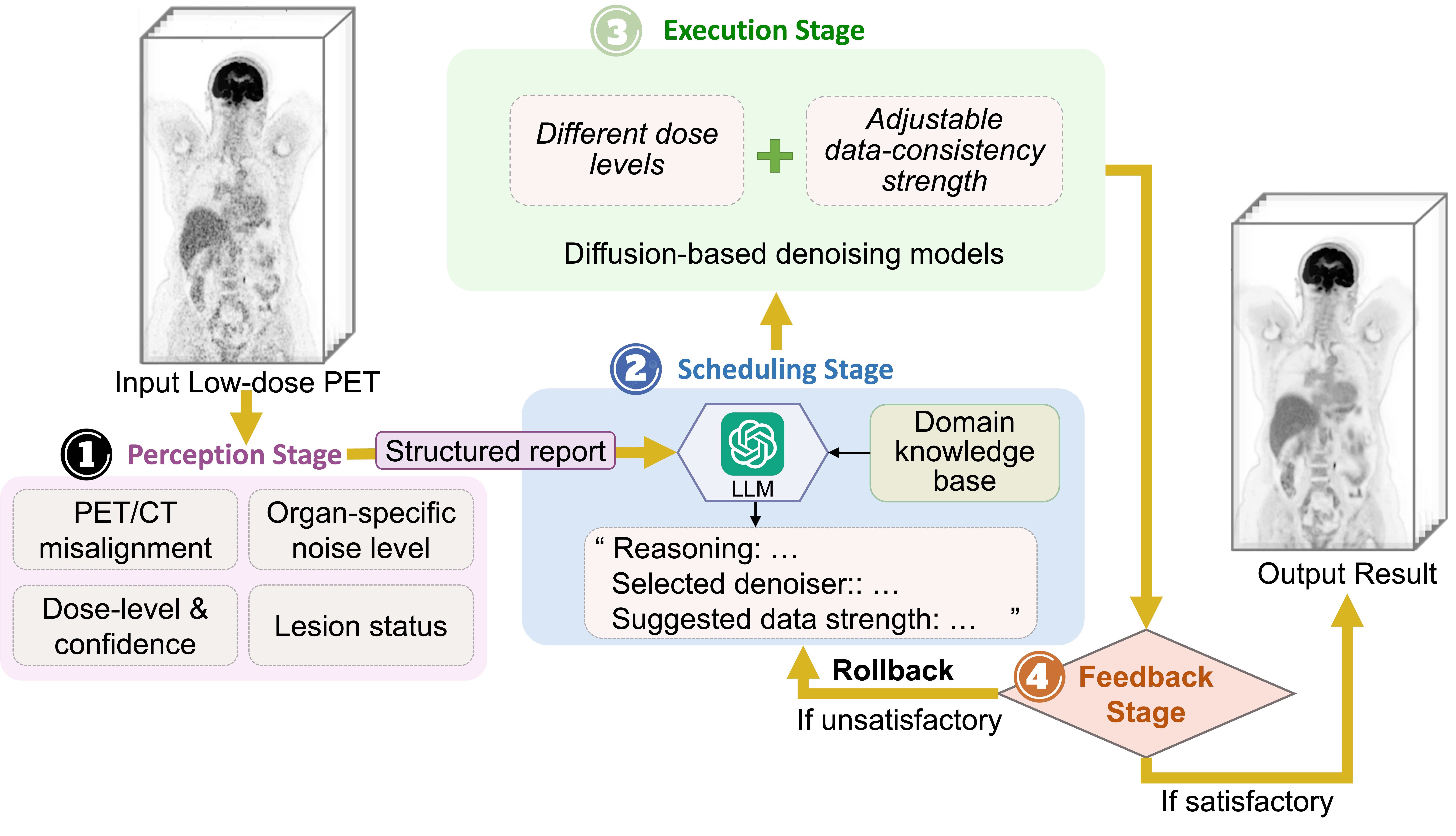}
\caption{Overview of the proposed closed-loop multi-agent framework for low-dose PET image denoising.} \label{overall_workflow}
\end{figure}

In this work, we propose an intelligent multi-agent framework for low-dose PET image denoising driven by VLM/LLM reasoning. As illustrated in Fig.~\ref{overall_workflow}, the proposed framework formulates PET denoising as a closed-loop process consisting of perception, scheduling, execution, and feedback. The system first performs multi-aspect perception: an anatomy-aware agent analyzes PET/CT consistency and identifies potential PET-CT mismatches; a dose-aware agent estimates the dose/count level from the input PET using a fine-tuned BiomedCLIP VLM; and a lesion-aware agent segments lesions and summarizes lesion status for downstream decision-making. Based on the structured perception report and a curated knowledge base derived from empirical PET denoising experience, an LLM-based scheduler selects an appropriate dose-specific conditional diffusion denoiser and corresponding inference parameter settings, including the strength of a data-consistency constraint. After denoising, a feedback agent re-evaluates organ-level noise reduction and checks quantitative stability, with explicit safeguards for lesion preservation. If the output does not meet predefined criteria, the system rolls back and reschedules with updated context. Together, these components provide a unified and automated solution for robust PET denoising across diverse dose/count levels.

\section{Related Work}

Agentic image restoration has recently emerged as a new paradigm for coordinating multiple specialized models within a unified image-processing workflow. In natural image restoration, methods such as RestoreAgent~\cite{chen2024restoreagent}, AgenticIR~\cite{zhu2025intelligent}, MAIR~\cite{jiang2026multi}, and 4KAgent~\cite{zuo20254kagent} decompose complex restoration into specialized stages for degradation perception, restoration planning, model or tool selection, execution, and potential reflection and rescheduling. These studies demonstrate the value of agentic coordination when an image contains heterogeneous degradations and requires different restoration tools/parameters. However, transferring this paradigm to quantitative medical imaging introduces additional requirements. Natural image restoration is commonly optimized according to perceptual quality or generic image fidelity, whereas PET processing must also preserve accurate quantitative uptake. A visually improved PET image can still be problematic if organ or lesion uptake is altered. PET restoration therefore requires domain-specific perception and quantitative safeguards, such as organ-level noise assessment and lesion-focused evaluation. In addition, general-purpose LLMs and VLMs do not inherently contain the fine-grained empirical knowledge required to select PET denoisers and tune their parameters under different imaging conditions.

Beyond general image restoration, agentic systems have also begun to appear in medical imaging, including image retrieval~\cite{vahistha2026agent}, radiology report generation~\cite{xiao2026multi}, collaborative image interpretation~\cite{zhang2025radagents}, and model development~\cite{tzanis2025maistro}. Most of these systems focus on diagnostic reasoning or language-centered clinical tasks rather than adaptive image restoration. AgentMRI~\cite{sajua2026agentmri} represents an important step toward agentic medical image processing by using a VLM to identify MRI degradations and route images to task-specific reconstruction, motion-correction, or denoising models. However, its current workflow relies on predefined decision logic and primarily performs degradation recognition followed by model selection. It does not explicitly evaluate the corrected image to guide subsequent strategy revision. For quantitative image processing, particularly PET denoising, a framework that connects multi-aspect image perception, model and parameter selection, quantitative verification, and failure-driven replanning remains underexplored.

\section{Methodology}

\subsection{Overview of the Multi-Agent Framework}

We propose an automated multi-agent system for low-dose PET denoising that follows a practical expert workflow: analyze the scan, select an appropriate denoising strategy, perform denoising, and verify the output quality. 

\begin{figure*}[t]
\centering
\includegraphics[width=0.8\textwidth]{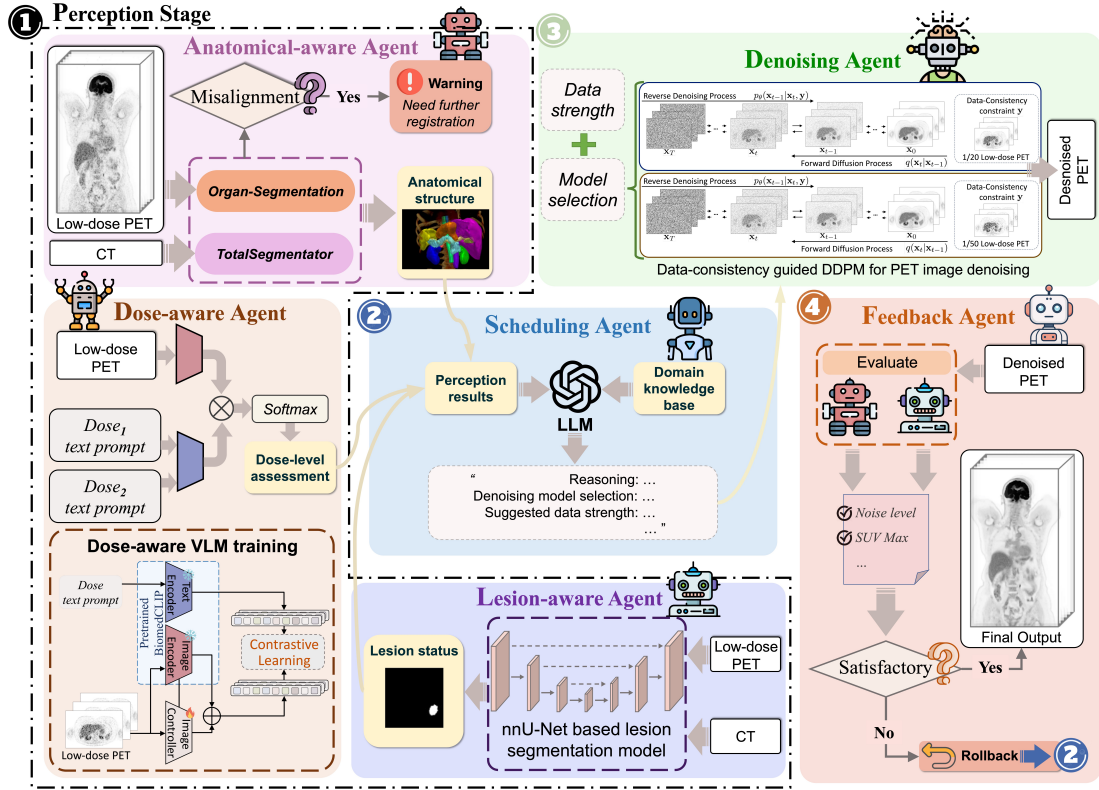}
\caption{Detailed architecture of the proposed multi-agent PET denoising system.} \label{agent_architecture}
\end{figure*}

The pipeline begins with a perception stage, where three specialist agents analyze the input scan from complementary perspectives. Specifically, an anatomy-aware agent checks PET/CT misalignment; a dose-aware agent estimates the noise condition of the PET scan; and a lesion-aware agent characterizes lesion presence and location. In the scheduling stage, an LLM integrates the perception outputs with distilled domain knowledge to select an appropriate denoiser and associated inference settings. In the execution stage, the selected denoising model is applied to generate the enhanced PET image. The system then enters a feedback stage that reassesses organ-level noise reduction and lesion-related quantification. If predefined criteria are not satisfied, the system triggers rollback and rescheduling. This design supports diverse low-dose conditions and reduces manual intervention in routine denoising and validation. Fig.~\ref{agent_architecture} illustrates the detailed architecture of the proposed multi-agent system.

\subsection{Perception for PET Image Understanding}

Reliable automation requires the system to recognize image quality issues and clinically relevant content that would typically be assessed by experts. We therefore implement a perception stage with three agents.

PET/CT misalignment can occur due to respiratory motion and patient motion. Such misregistration may affect lesion localization and quality assessment. The anatomy-aware agent generates organ masks from both modalities~\cite{salimi2025deep,wasserthal2023totalsegmentator} and compares organ locations and overlaps, with emphasis on motion-sensitive organs such as the lung and liver. When organ disagreement indicates potential PET/CT misalignment, the agent issues a warning suggesting additional registration. In addition, organ-specific quality assessment is enabled by the segmentation masks, allowing the agent to compute organ-level noise statistics that guide denoiser selection and serve as reference measurements for post-denoising evaluation.

To support dose-adaptive denoising, we develop a dose-aware agent. Specifically, we fine-tune BiomedCLIP~\cite{zhang2023biomedclip}, a large-scale pretrained biomedical VLM, to estimate the image-derived dose condition. Each dose level is represented by a text prompt, and the model predicts the level whose prompt embedding is most similar to the image embedding in the shared representation space. To preserve general biomedical semantics, we freeze the original text and image encoders of BiomedCLIP and introduce a lightweight image-side controller initialized from the pretrained image encoder. The controller is optimized using contrastive learning so that embeddings of low-dose PET images align with their corresponding dose prompts. For a 3D PET scan, predictions are first produced at the axial-slice level and then aggregated to yield a subject-level estimate and confidence score.
Importantly, this agent is designed to characterize the image-derived noise condition rather than simply recover the nominal acquisition label. Reconstructed PET noise depends not only on the dose level, but also on patient weight, scan duration, and scanner and reconstruction settings. Therefore, acquisition metadata alone does not fully characterize the noise observed in the reconstructed PET image. The VLM directly analyzes image appearance and provides a confidence score together with the predicted dose condition. This information is subsequently used for dose-specific model and parameter selection.

The lesion-aware agent performs whole-body lesion segmentation and summarizes lesion status for scheduling and feedback validation. We train a 3D nnU-Net for PET lesion segmentation, with a ResNet encoder and a larger patch size (192$\times$192$\times$192) to increase the receptive field and improve robustness for whole-body coverage. The agent outputs a lesion mask and basic lesion descriptors, including lesion SUV\textsubscript{max}. 

Outputs from the three perception agents are converted into a compact, structured report for the scheduler. The report includes potential PET/CT alignment warnings, organ-level noise statistics, estimated dose level with confidence, and lesion summaries. 

\begin{figure*}[htb]
\centering
\includegraphics[width=0.9\textwidth]{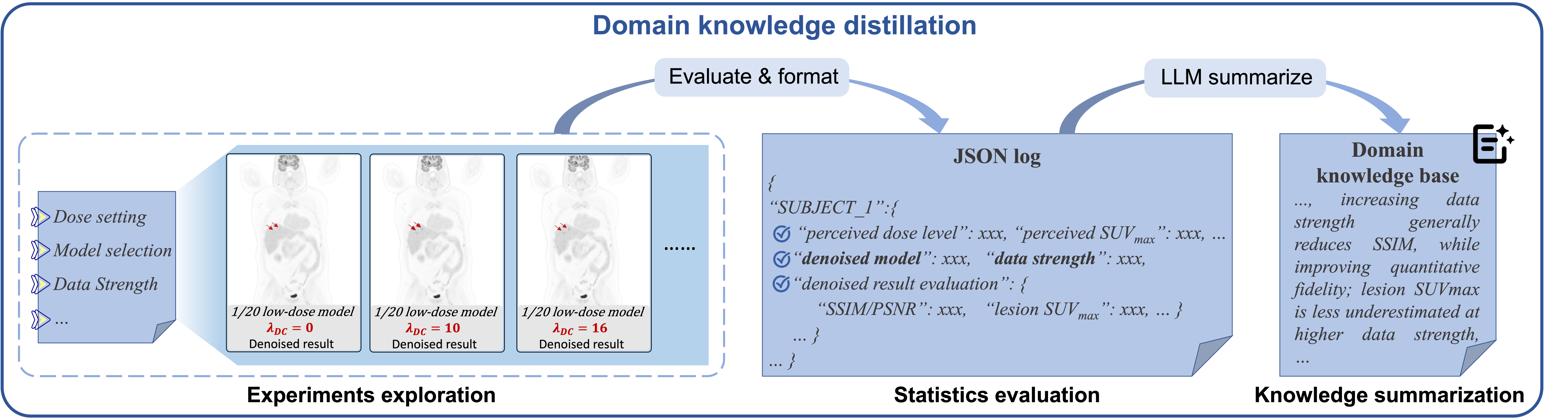}
\caption{Construction of the PET denoising domain knowledge base.} \label{knowledge_distillation}
\end{figure*}

\subsection{LLM-based PET Image Denoising Scheduling and Closed-loop Feedback}

Scheduling requires contextual reasoning across multiple statistics, including dose level, organ noise, lesion status, and alignment conditions. Fixed rule-based systems are difficult to maintain under such real-world variability. We therefore employ an LLM as the scheduling hub to integrate these heterogeneous perception outputs with PET-specific domain knowledge and information from previous attempts. This process mimics how PET experts assess the image condition, select a denoising strategy, review the result, and revise the strategy when necessary.
In this feasibility study, we use GPT-4 (gpt-4o-mini) as the reasoning engine.

The LLM is not guaranteed to possess PET denoising domain knowledge or to reliably select among specialized models. To address this, we provide a curated domain knowledge base derived from empirical exploration, including common failure modes (e.g., over-smoothing and lesion disappearing), recommended model choices for different dose levels, and suitable ranges for key inference settings. 
The construction of this knowledge base is illustrated in Fig.~\ref{knowledge_distillation}. Different denoising strategies were first explored on the validation cases by varying the available denoisers and the data-consistency strength $\lambda_{\mathrm{dc}}$ (Sec.~\ref{sec:ddpm}). For each experiment, the corresponding perceptual information, denoising settings, and evaluation results were recorded. The accumulated records were then summarized by the LLM into PET-specific denoising knowledge, which describes preferred models and ranges of $\lambda_{\mathrm{dc}}$ under different perception conditions and how $\lambda_{\mathrm{dc}}$ affects the balance between noise suppression and uptake preservation. Given the structured perception report and retrieved domain knowledge, the LLM generates a structured JSON plan specifying the dose-specific conditional diffusion model and the recommended data-consistency strength.

After denoising, a feedback agent assesses organ-level noise reduction and quantitative stability by comparing organ bias and lesion SUV metrics before and after denoising. If the output fails to meet predefined criteria (e.g., insufficient noise reduction or excessive lesion SUV change), the system rolls back to the pre-denoising state. The previous denoising plan and its feedback outcome are then included in the next scheduling round, allowing the LLM to revise the strategy according to the observed failure. This closed-loop mechanism improves robustness and reduces the need for manual trial-and-error.

\subsection{Data-Consistency Guided DDPM for Low-Dose PET Denoising}
\label{sec:ddpm}

We train dose-specific conditional Denoising Diffusion Probabilistic Models (DDPMs) using paired low-dose and normal-dose PET data. Let $y$ denote the low-dose PET input and $x_0$ denote the corresponding normal-dose PET target. We use the standard forward diffusion process
\begin{align}
q(\mathbf{x}_{1:T} | \mathbf{x}_0) &= \prod_{t=1}^T q(\mathbf{x}_t | \mathbf{x}_{t-1} ), \\
q(\mathbf{x}_t|\mathbf{x}_{t-1}) &= \mathcal{N}(\mathbf{x}_t;\sqrt{1-\beta_t}\mathbf{x}_{t-1},\beta_t \mathbf{I}), 
\end{align}
which implies the forward process can be further expressed as
\begin{align}
  q(\mathbf{x}_t|\mathbf{x}_0) &= \mathcal{N}(\mathbf{x}_t; \sqrt{\bar\alpha_t}\mathbf{x}_0, (1-\bar\alpha_t)\mathbf{I}), \\
 \bar{\alpha}_t&=\prod_{s=1}^t (1-\beta_s).
\end{align}
A conditional denoiser $\epsilon_\theta$ predicts the added noise given the noisy sample, timestep $t$, and the conditioning low-dose input. We optimize the standard noise-prediction objective by
\begin{align}
\mathcal{L}_{\mathrm{DDPM}}
=\mathbb{E}_{t, \mathbf{x}_0, \boldsymbol{\epsilon}}\!\left[{ \left\| \boldsymbol{\epsilon} - \boldsymbol{\epsilon}_\theta(\mathbf{x}_t,  \mathbf{y}, t) \right\|^2}\right].
\end{align}
In practice, purely prior-driven generation can introduce bias in PET, and small lesions may be overly smoothed. To address this, during sampling we add a data-consistency term that pulls the prediction toward the measured low-dose PET image as~\cite{gong2023pet}
\begin{equation}
\begin{aligned}
\mathbf{x}_0&=\frac{1}{\sqrt{\bar{\alpha}_t}}
\left(\mathbf{x}_t-\sqrt{1-\bar{\alpha}_t}\,\boldsymbol{\epsilon}_\theta(\mathbf{x}_t,  \mathbf{y}, t)\right) \\ 
&\quad + \eta_t\,\lambda_{\mathrm{dc}}\,(\mathbf{y}-\mathbf{x}_t),
\end{aligned}
\label{eq:dc_update}
\end{equation}
where $t$ is the time step, $\lambda_{\mathrm{dc}}$ is the data-consistency strength, and $\eta_t$ is a step-size schedule. 
Equation~\eqref{eq:dc_update} enables a controlled trade-off between the diffusion prior and measurement fidelity. The LLM scheduler selects $\lambda_{\mathrm{dc}}$ based on dose estimation, organ-level noise statistics, and lesion presence. The feedback loop adjusts it when quantitative checks indicate under- or over-regularization.

\section{Experiments}

\begin{figure*}[htb]
\centering
\includegraphics[width=0.8\textwidth]{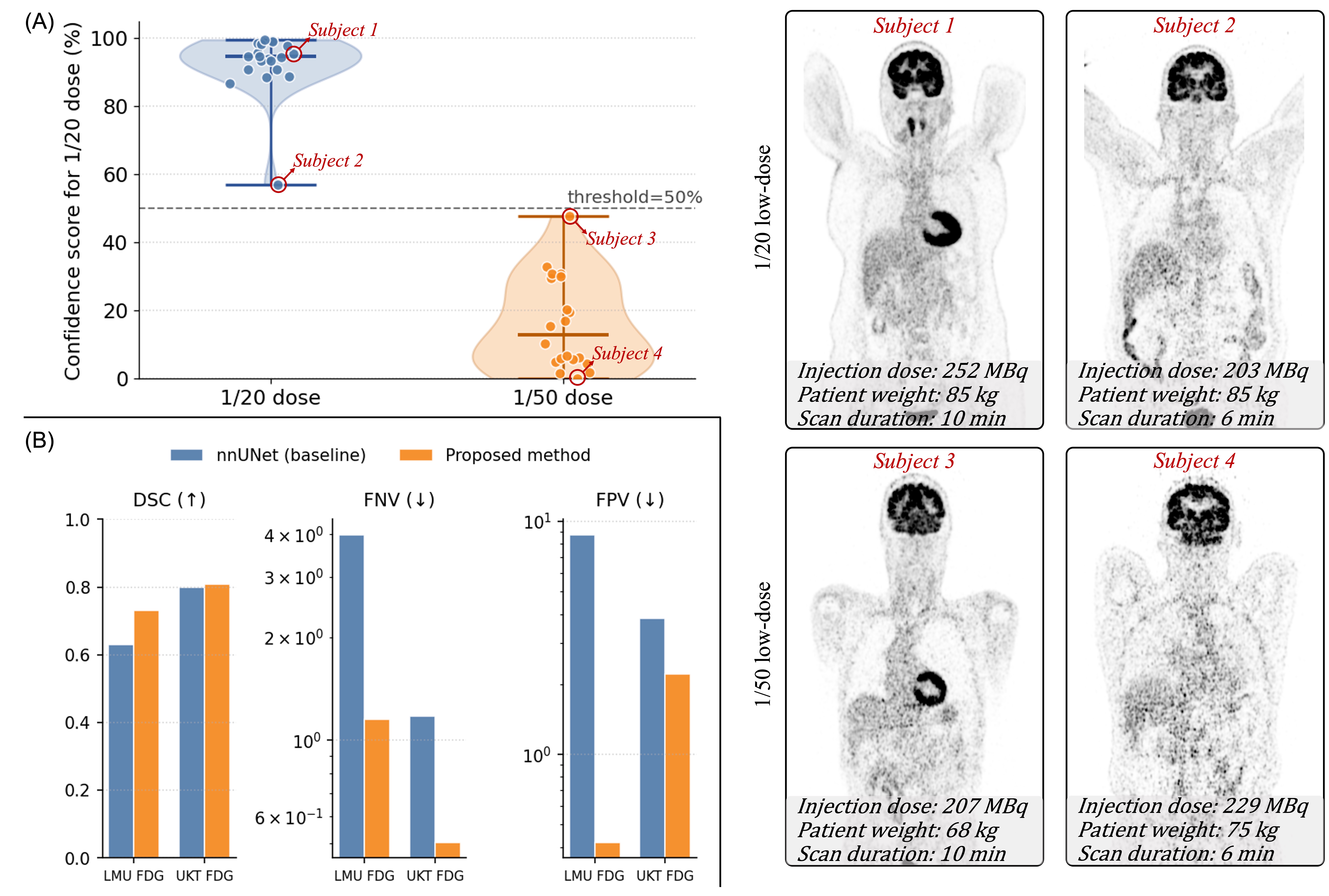}
\caption{Evaluation of dose-aware perception and lesion-aware segmentation. (A) Violin plots of the dose-level confidence predicted by the fine-tuned BiomedCLIP. A 50\% threshold is used to distinguish the $1/20$- and $1/50$-dose predictions. Representative subjects with relatively high and low confidence scores within each dose group are illustrated, with corresponding PET images and acquisition-related information shown on the right. (B) Lesion segmentation performance of the lesion-aware agent on the AutoPET-IV FDG test subsets (LMU and UKT), compared with the challenge baseline, reported with foreground DSC, FNV, and FPV.} \label{fig4}
\end{figure*}

\subsection{Datasets }
BiomedCLIP fine-tuning (dose estimation) and the DDPM models were developed using a Siemens Biograph Vision Quadra PET/CT dataset with two low-dose settings ($1/20$ and $1/50$ of full counts). The $^{18}$F-FDG PET images were reconstructed using OSEM ($4$ iterations, $5$ subsets) with time-of-flight (TOF) and point spread function (PSF) modeling, followed by a $2\ \mathrm{mm}$ Gaussian filter~\cite{alberts2021clinical}. The voxel spacing was $1.65 \times 1.65 \times 1.65~\mathrm{mm}^3$. All data were acquired in list mode, and the 1/20 and 1/50 low-dose scans were generated by list-mode rebinning that subsampled a corresponding fraction of the original full-dose counts. Image intensity was represented using SUV. We used $115$ whole-body scans, which were split into $90/5/20$ for training/validation/testing, respectively. Background regions were cropped for computational efficiency, yielding volumes of $192 \times 288 \times 520$ (coronal $\times$ sagittal $\times$ axial). For denoising, two dose-specific conditional DDPMs were trained separately: one using paired $1/20$ low-dose and normal-dose PET, and the other using paired $1/50$ low-dose and normal-dose PET. The lesion-aware agent employed a 3D nnU-Net trained on the MICCAI 2025 AutoPET-IV dataset [training: $1{,}014$ FDG and $597$ PSMA subjects; test: $100$ subjects with $50$ FDG LMU (from LMU University Hospital Munich clinical center) and $50$ FDG UKT (from University Hospital Tübingen clinical center)].

\subsection{Implementation Details}
For training and testing of DDPM models,  two adjacent axial slices were jointly fed into the network to mitigate axial artifacts. All models were implemented in PyTorch with a learning rate of $1 \times 10^{-4}$ and mixed-precision training for computational efficiency. In both the forward and reverse diffusion processes, the time step $T$ was set to $1000$, with the variance schedule $\beta_t$ linearly increasing from $1 \times 10^{-4}$ to $0.02$. The total training batch size was set to $48$, and distributed training was performed on $6$ NVIDIA B200 GPUs. Training required approximately two days, and the average inference time was $10$ minutes per dataset. For comparison, UNet, GAN, and DDPM baselines were trained on the combined $1/20$ and $1/50$ low-dose datasets. In contrast to the proposed framework, the DDPM baseline used a single model for both dose levels and did not employ dose-specific model selection or adaptive data-consistency guidance.

For lesion-segmentation evaluation, we calculated the foreground Dice similarity coefficient (DSC). Detection errors were further quantified using false positive volume (FPV), defined as the total volume of predicted connected components not overlapping any ground-truth lesion, and false negative volume (FNV), defined as the total volume of ground-truth lesion components missed by the prediction. For denoising evaluation, Peak Signal-to-Noise Ratio (PSNR) and Structural Similarity Index Measure (SSIM) were computed using normal-dose PET as reference. Statistical significance of PSNR/SSIM differences between methods was assessed using Wilcoxon signed-rank tests.

\begin{figure*}[htb!]
\centering
\includegraphics[width=0.8\textwidth]{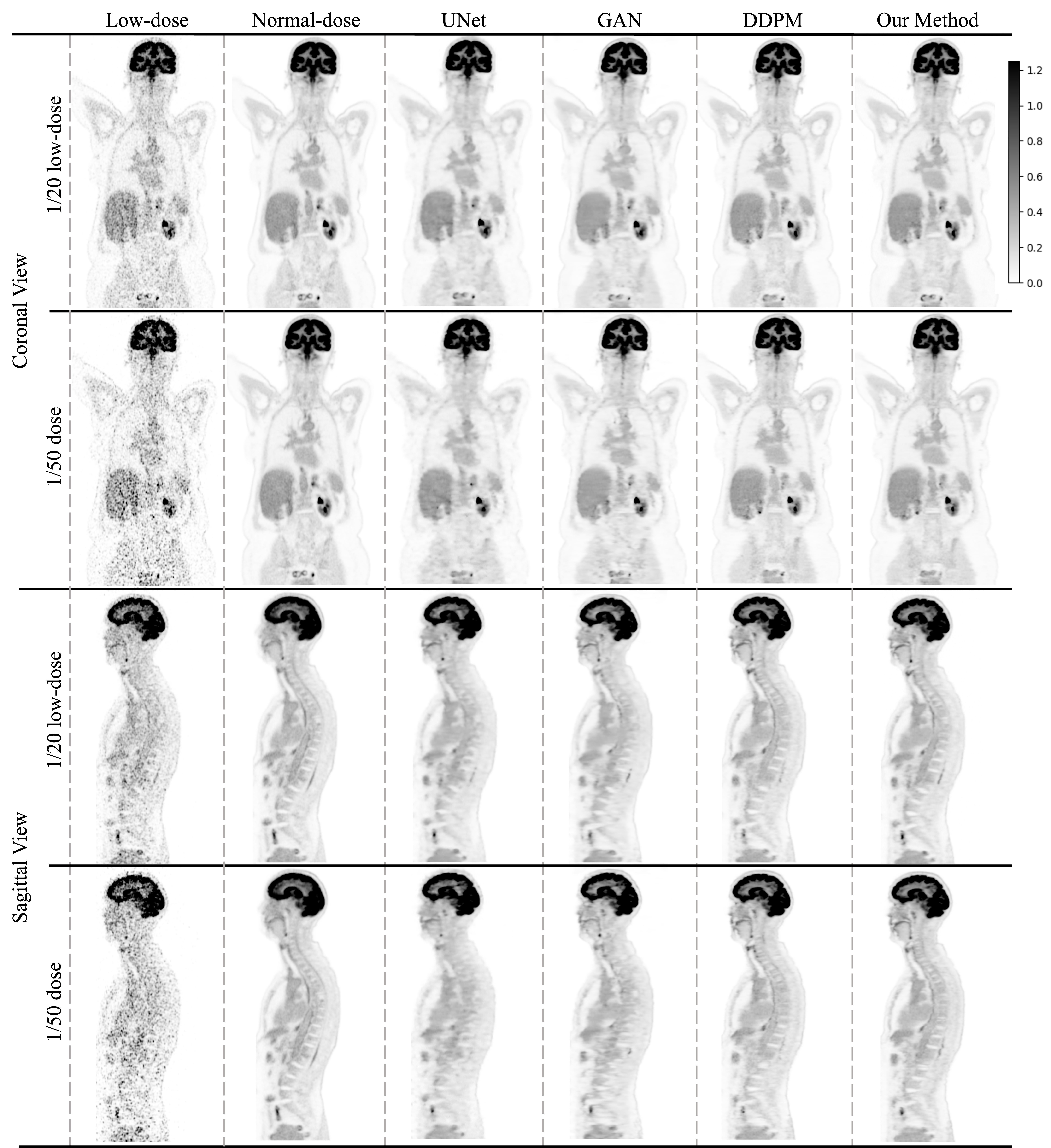}
\caption{Sagittal views of a Siemens Biograph Vision Quadra $^{18}$F-FDG test data with $1/20$ and $1/50$ low-dose PET images denoised using different methods, alongside the corresponding normal-dose PET image.} \label{fig5}
\end{figure*}

\section{Results}

\subsection{Performance of the Perception Modules}

As shown in Fig.~\ref{fig4}(A), the dose-aware agent showed clear separation between the $1/20$- and $1/50$- dose groups using a 50\% confidence threshold. At the same time, noticeable variations were observed within each dose group. For example, Subjects 1 and 2 were both generated at the $1/20$ dose level but showed differences in image noise and substantially different confidence scores. A similar variation was observed between Subjects 3 and 4 within the $1/50$ group. Their injection dose, patient weight, and scan duration also differed, illustrating that the nominal dose group alone does not fully characterize the reconstructed PET noise condition. These results support the use of the dose-aware agent to provide image-derived information for subsequent denoising scheduling rather than relying solely on the predefined dose level.

\vspace{-4mm}
\begin{figure}[htb]
\centering
\includegraphics[width=0.5\textwidth]{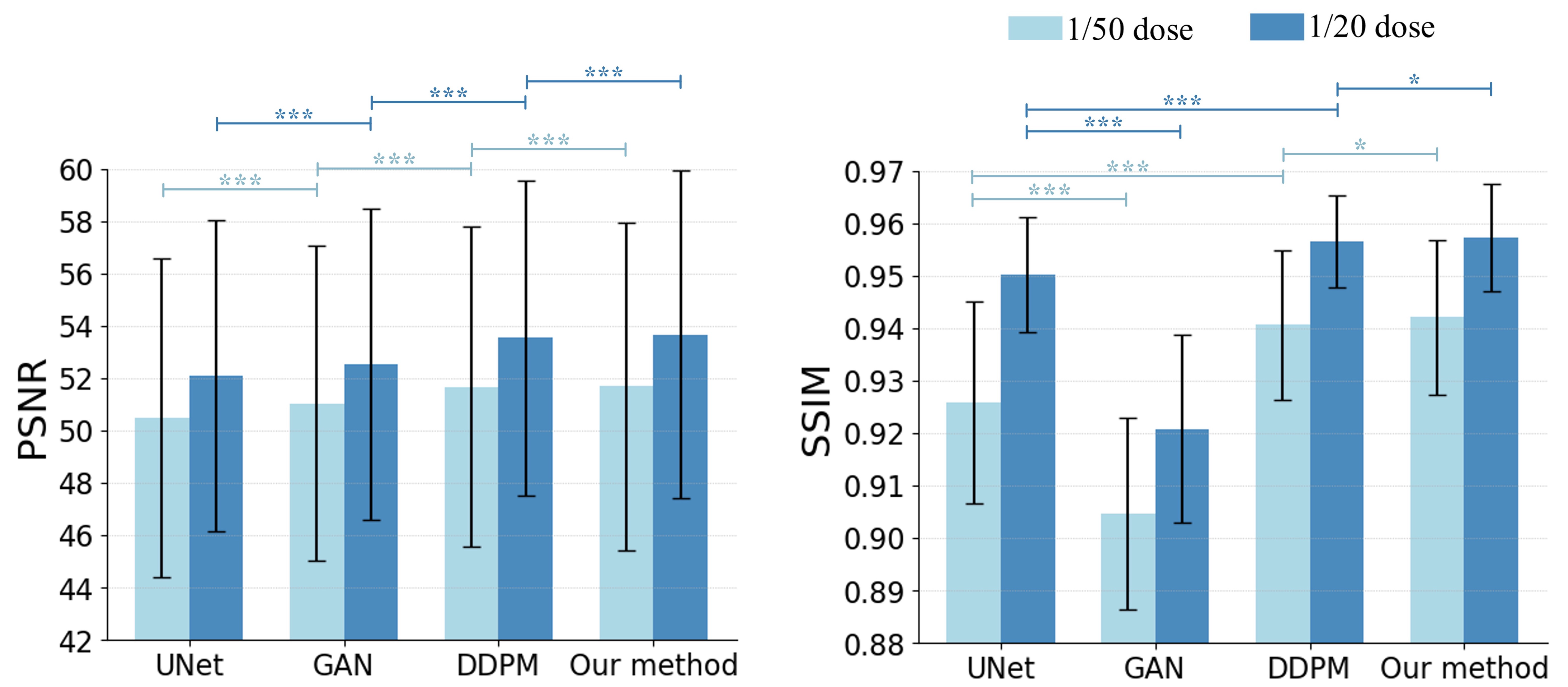}
\caption{Quantitative evaluation of overall PET denoising performance (PSNR and SSIM) on 20 test cases for each of the $1/20$- and $1/50$ low-dose settings. Statistical significance is indicated by * ($p\textless0.05$) and *** ($p\textless0.001$).} \label{fig6}
\end{figure}


Fig.~\ref{fig4}(B) evaluates the lesion-aware agent on the AutoPET-IV FDG test subsets from LMU and UKT. Both the conventional nnU-Net baseline and our modified 3D nnU-Net achieved effective lesion segmentation, while the proposed model further improved foreground DSC and reduced FNV and FPV on both clinical centers' test subsets. These results demonstrate that the lesion-aware agent can provide reliable lesion information for subsequent scheduling and feedback evaluation.

\begin{figure*}[htb]
\centering
\includegraphics[width=0.8\textwidth]{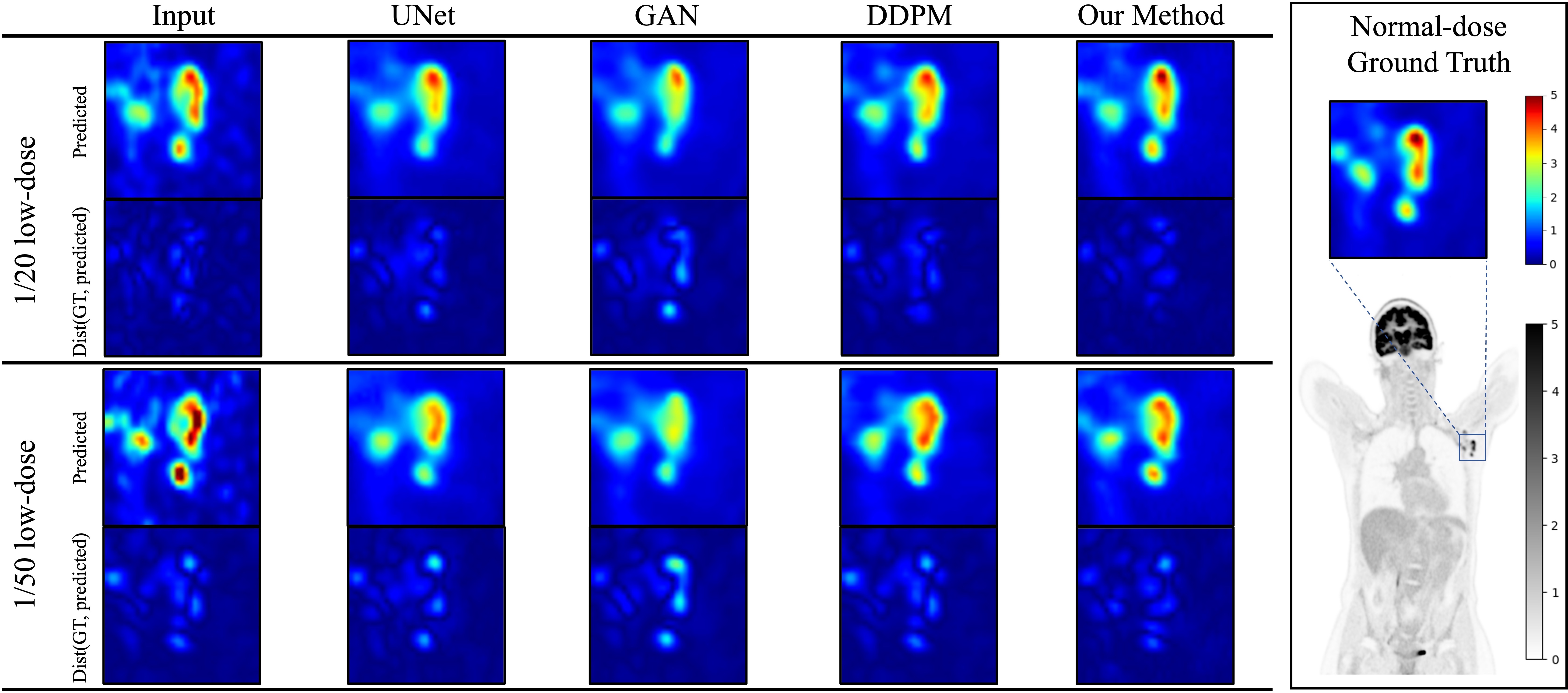}
\caption{Zoomed-in lesion region and L$^{2}$ norm distance images between the predicted and GT images of a Siemens Biograph Vision Quadra $^{18}$F-FDG test data, alongside the corresponding normal-dose PET image.} \label{fig7}
\end{figure*}

\subsection{End-to-End Denoising Performance of the Multi-Agent Framework}

Fig.~\ref{fig5} compares the qualitative denoising results under the $1/20$- and $1/50$-dose settings. The low-dose inputs exhibited substantial noise and reduced signal-to-noise ratio. All methods improved image quality, while UNet and GAN showed noticeable over-smoothing. The DDPM baseline better preserved anatomical structures, demonstrating the effectiveness of diffusion model-based PET denoising. The proposed multi-agent framework, which selects a dose-specific DDPM and adaptively determines the data-consistency strength, further reduced residual artifacts and produced image appearances closer to the normal-dose reference. Quantitative results in Fig.~\ref{fig6} show that the proposed method achieved the highest PSNR and SSIM at both dose levels, with statistically significant improvements over the reference methods.

The lesion-focused comparisons in Fig.~\ref{fig7} further demonstrate the benefit of the complete framework. Although the DDPM baseline provided effective overall denoising, local differences remained around lesion regions. In contrast, the proposed framework combines dose-specific denoising with an adaptive data-consistency constraint, resulting in better preservation of lesion uptake patterns and lower L$^{2}$ distance errors. This improvement is particularly important for PET, where excessive denoising may suppress small or high-uptake structures. By combining dose-aware model selection with adjustable data-consistency strength and post-denoising feedback, the proposed framework achieved improved whole-body image quality while maintaining better lesion-region fidelity.

\section{Discussion}
The dose-aware agent provides an image-derived assessment that is more informative for denoising scheduling than directly using the nominal dose level from acquisition metadata. The same injected activity does not necessarily produce the same image noise across scanners or imaging protocols, since reconstructed PET quality is also affected by scanner sensitivity, acquisition duration, and reconstruction settings. A metadata-based strategy would therefore require protocol-specific rules to map acquisition settings to an appropriate denoiser. In contrast, our VLM directly evaluates the reconstructed PET appearance and assigns it to an image-derived dose condition. This formulation provides a practical way to group scans with similar noise characteristics even when they originate from different acquisition settings, reducing the need to define separate decision rules for each protocol. In the current study, two low-count conditions are used to demonstrate the feasibility of this image-based assessment. The same formulation can be extended to additional noise conditions when corresponding training data and denoising models are available.

LLM hallucination remains an important concern for agentic medical imaging systems. In our framework, this risk is reduced through several constraints. First, the distilled domain knowledge bounds the LLM's $\lambda_{\mathrm{dc}}$ and model choices to safe ranges, blocking out-of-scope plans at the source. Second, the feedback agent evaluates organ noise/bias and lesion SUV\textsubscript{max}, and triggers rollback when these fall outside expected ranges. Finally, data-consistency guidance restricts the diffusion output using information from the measured low-dose PET image. These mechanisms reduce the likelihood that an unsupported LLM decision is directly accepted as the final result. Nevertheless, they do not completely remove the uncertainty associated with LLM output. Future studies will evaluate scheduling consistency across repeated runs and different language models.

An important feature of the proposed framework is its modular organization, which allows individual components to be replaced or extended without redesigning the entire workflow. In this study, our main goal is to validate the closed-loop agentic framework rather than to develop a new denoising backbone. The current DDPM-based execution module can therefore be replaced by more advanced PET denoisers, which may further improve image quality and computational efficiency. The same flexibility applies to the perception and feedback modules, allowing new image-quality assessments or quantitative criteria to be incorporated when needed. This modular design also facilitates extension beyond image denoising, and future work will investigate its application to diffusion-based PET reconstruction. In addition, the current computational burden is mainly associated with iterative DDPM sampling rather than the agentic workflow itself. Future implementations may incorporate flow-matching or other accelerated generative models to reduce inference time while retaining the same perception, scheduling, and feedback framework.

\section{Conclusion}
This VLM/LLM-enabled multi-agent framework delivers robust low-dose PET denoising across dose levels while preserving lesion details. By automating image quality/lesion perception, denoising scheduling, and closed-loop validation with rollback, it can reduce reliance on manual review and streamline the denoising workflow.
\section{Acknowledgement}
All authors declare that they have no known conflicts of interest in terms of competing financial interests or personal relationships that could have an influence or are relevant to the work reported in this paper.

\bibliographystyle{IEEEtran}
\bibliography{mybibliography}

\end{document}